\documentclass[aps,ams,preprint,showpacs,amsmath,amssymb,superscriptaddress]{revtex4-1} 
\usepackage{graphicx}
\usepackage{multirow}
\usepackage{hyperref}

\usepackage{color}
\usepackage{soul}

\begin{document}


%
%

\title{Lateral Casimir force between deeply corrugated dielectric and metallic plates}
\author{ARASH AZARI}
\affiliation{School of Computational Sciences, Korea Institute for Advanced Study, Seoul 130-722, Republic of Korea\\
Department of Physics and Astronomy, University of Sheffield, Sheffield S3 7RH, UK}

\author{MIRFAEZ MIRI }
\affiliation{Department of Physics, University of Tehran, P.O. Box 14395-547, Tehran, Iran\\
School of Physics, Institute for Research in Fundamental Sciences, (IPM) Tehran 19395-5531, Iran\\
miri@iasbs.ac.ir}

\received{October 2011}

\begin{abstract}
We study the lateral Casimir force between corrugated dielectric plates.
We use the dielectric contrast perturbation theory [R. Golestanian, {\it Phys. Rev. Lett.} {\bf 95}, 230601, (2005)], which allows us to consider arbitrary deformations with large amplitudes. We consider sinusoidal, rectangular, and sawtooth corrugations, for samples made of silicon and gold.
We use the plasma and Drude-Lorentz models for the permittivity of gold and silicon, respectively.
For these geometries and materials, the lateral Casimir force is {\it not} a sinusoidal function of the relative lateral
displacement of plates when the gap between the plates in comparable with the depth of the corrugations.
Our results facilitate the design of miniaturized devices based on lateral Casimir forces.
\end{abstract}

\maketitle
\section{Introduction}

The Casimir force\ \cite{Casimir48} between macroscopic objects originates from the quantum fluctuations of the electromagnetic vacuum.
It is known that the Casimir force strongly depends on the {material properties} and {geometry} of the objects\ \cite{pala,Klim-etal-RMP09,RG}.
For example, two perfectly metallic plates immersed in the vacuum experience the attractive Casimir force,
but a metallic plate and a purely magnetic one experience the repulsive force\ \cite{Boyer}.
Recently, Munday {\it et al.} observed a long-range repulsive force between gold and
silica particles that are separated by bromobenzene \cite{Capasso-re}.

It has been proposed theoretically\ \cite{lateral} and verified experimentally\ \cite{chen,chiu} that two corrugated plates experience {lateral} Casimir force.
Recently it has been suggested that the lateral Casimir force may intermesh
separate parts of nanomachines\ \cite{2007,AFR,someAFR}.
If the corrugated metallic plates are at a distance from each other that is considerably larger than the corrugation amplitudes, then the lateral force is a sinusoidal function of the relative lateral shift.
To better understand the possibilities that such nanomachines could bring about, one must consider realistic material properties and large corrugation amplitudes. We note that with the recent developments
in UV lithography and dry-etching techniques, the fabrication of such Casimir nanomachines
is not out of reach\ \cite{ema}.

Computing exact Casimir forces via standard numerical electromagnetism techniques is described\ \cite{exact}.
The scattering theory\ \cite{njp} also includes both geometry effects and the optical properties of materials.
For rectangular corrugations on ideal metallic plates, a nonperturbative approach to Casimir interaction is proposed\ \cite{PRA62101EMIG}.
To treat shallow corrugations, one expands the reflection operators in terms of corrugation amplitudes. In case of sinusoidally corrugated plates, described by perfect metal\ \cite{Emig2001} or plasma model\ \cite{over1}, the lateral force is calculated to second order in the deformation amplitude.

Indeed, {\it approximate} methods are quite useful in early stages of nanomachine design.
In case of shallow corrugations, the Casimir force can be estimated using the proximity force approximation (PFA)\ \cite{over1}. For metallic plates with sawtooth corrugation pairwise summation (PWS) is applied\ \cite{PRA44103LATERAL}. Here we use
dielectric contrast perturbation theory (DCP)\ \cite{Ramin00,Arash}, which allows us to consider arbitrary deformations with large amplitudes. We consider three different examples of sinusoidal, rectangular, and sawtooth corrugations. We demonstrate the results for two types of
silicon and gold plates. We use the plasma and Drude-Lorentz models for the permittivity of gold and silicon, respectively.

\section{Theoretical Formulation}

\begin{figure*}[t]
\includegraphics[width=.31\columnwidth]{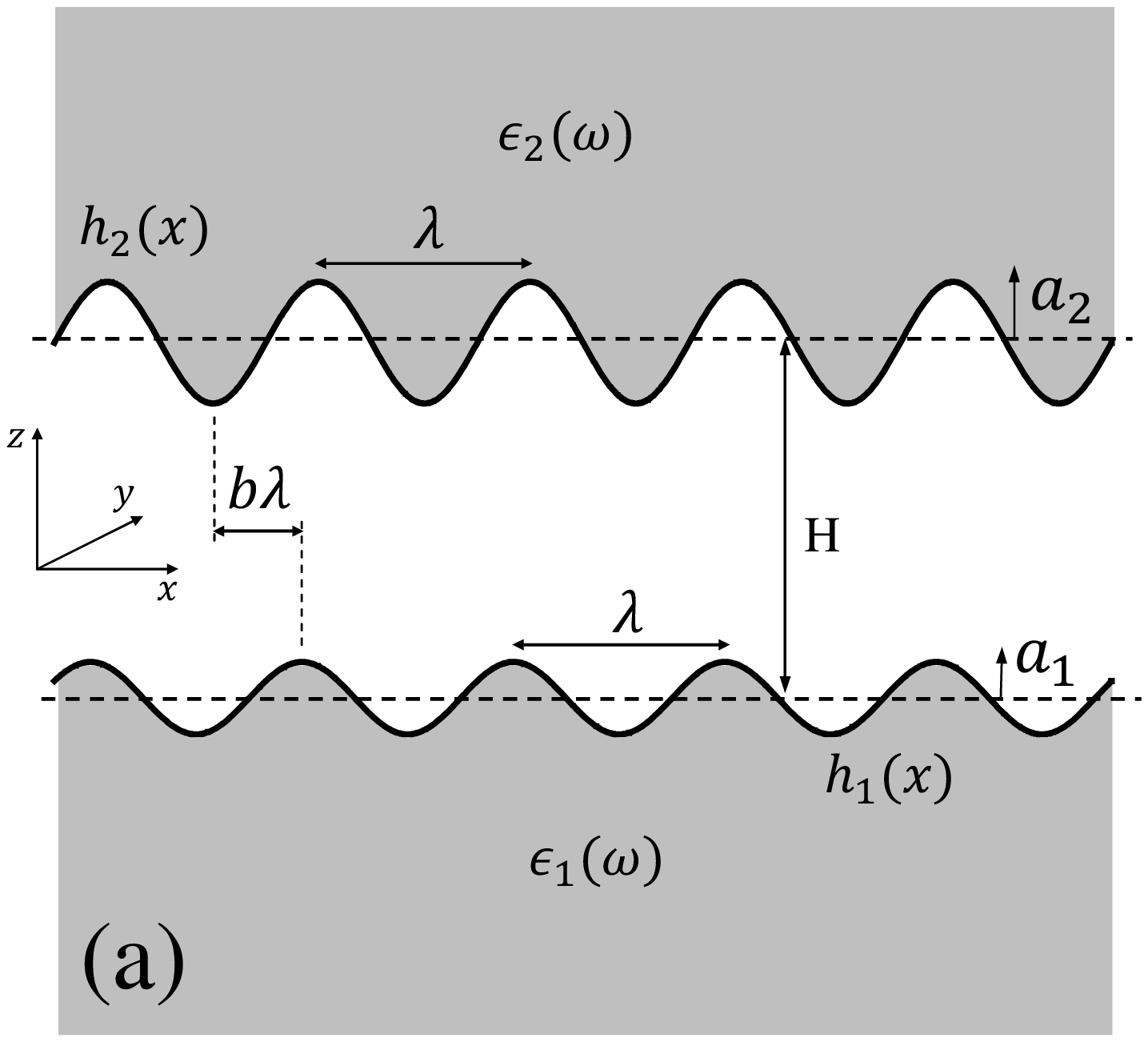}
\includegraphics[width=.31\columnwidth]{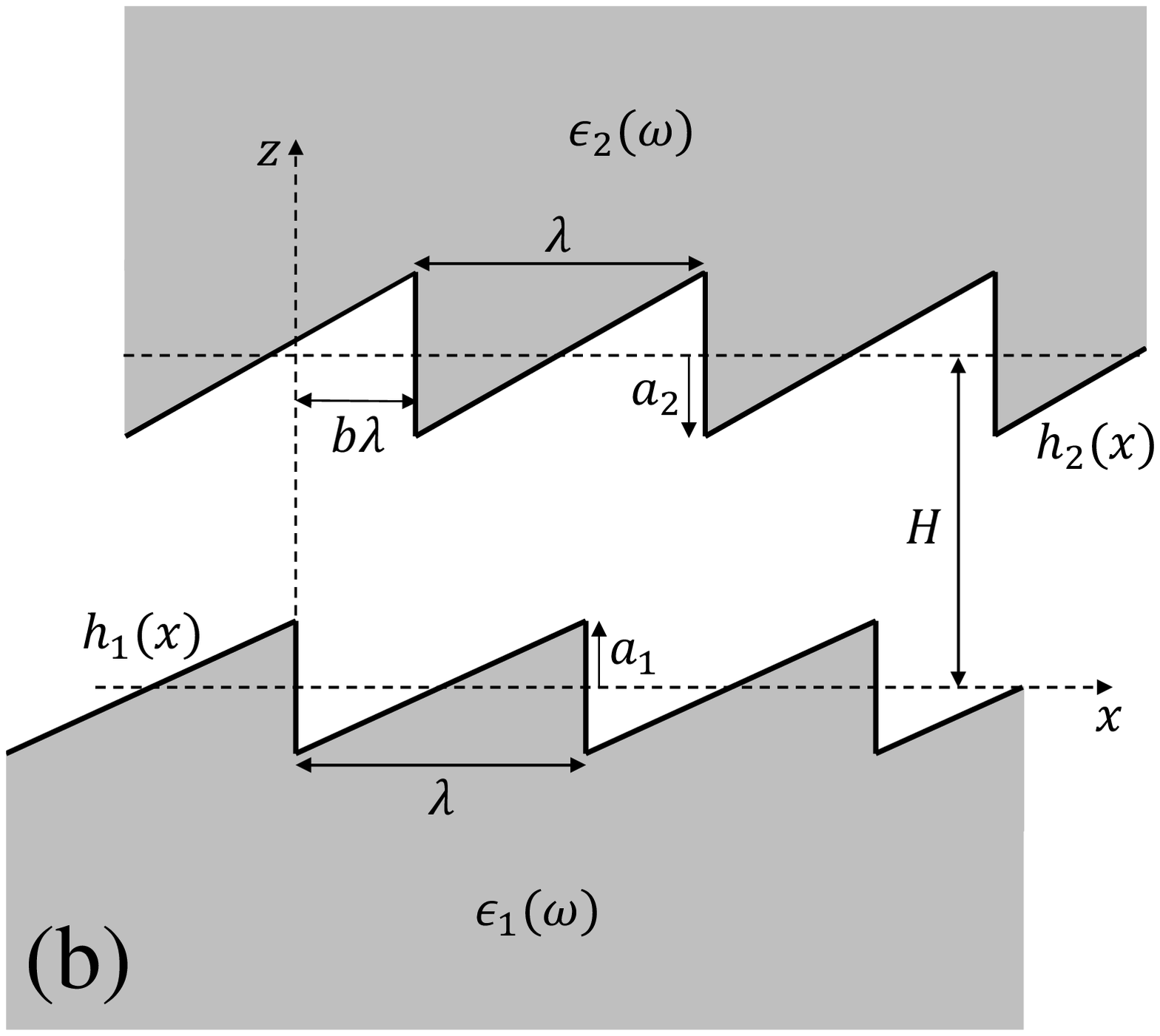}
\includegraphics[width=.35\columnwidth]{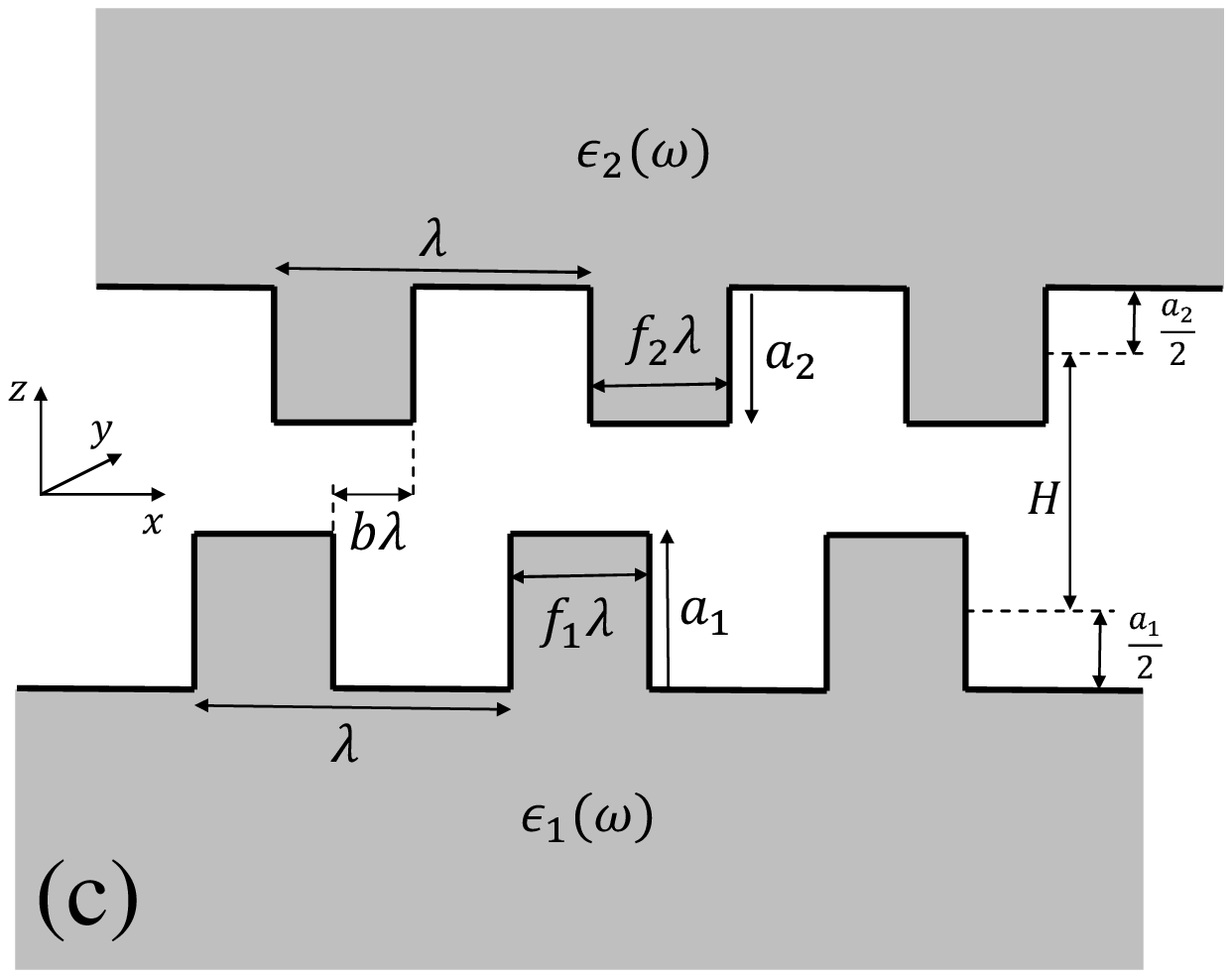}
\caption{Two corrugated plates with surface profiles $h_{1}({x})$ and $h_{2}({x})$.
$H$, $\lambda$, $b \lambda $, $a_{1}$ and $a_{2}$ denote the mean separation, corrugation wavelength, relative lateral displacement,
and corrugation amplitudes, respectively. Left) sinusoidal corrugations. Middle) sawtooth corrugations. Right) rectangular corrugations characterized with dimensionless
parameters $f_1$ and $f_2$.
} \label{fig:sifig}
\end{figure*}

We consider two semi-infinite dielectric bodies placed at mean separation $H$, see Fig. \ref{fig:sifig}. The height functions $h_{1}({\bf x})$ and
$h_{2}({\bf x})$ characterize surface profile of the plates, where ${\bf x}=(x,y)$ denotes the lateral coordinates. The (imaginary) frequency dependent dielectric function of the system is
\begin{equation}
\epsilon \left(i \zeta,{\bf r}\right)=\left\{\begin{array}{ll}
\epsilon_2\left(i \zeta\right) & \; H + h_2\left( {\bf x} \right) \leq z < +\infty,   \\
\\
1 & \; h_1\left({\bf x} \right) < z < H + h_2\left( {\bf x} \right),   \\
\\
\epsilon_1\left(i \zeta\right) & \; -\infty < z \leq h_1\left( {\bf x}\right).
\end{array} \right. \label{epsilon-profile}
\end{equation}
We use the Clausius-Mossotti resummation of the perturbative expansion in terms of the dielectric contrast
$\delta \epsilon\left(i \zeta\right)=\epsilon\left(i \zeta\right)-1$, which yields the Casimir energy of the system to the lowest order as\ \cite{Ramin00,Arash}
\begin{eqnarray}
E_{C} \! \!&=&\!-\!\frac{\hbar}{4 \pi^{2} c^{2}} \! \! \int_0^\infty \!\!d \zeta\; \!\zeta^{2}\, \overline{\delta\epsilon_{1}}(i \zeta)\,
\overline{\delta\epsilon_{2}}(i \zeta)
  \!\int \!\! d^{2} {\bf x}\,d^{2} {\bf x}^{\prime}\, {\mathcal M} \left({\bf x}\!\!-\!\!{\bf x}^{\prime},H\!\!+\!\!h_{2}
\left({\bf x}\right)\!\!-\!\!h_{1}({\bf x}^{\prime}),\zeta\right). \label{FinalEnergy4}
\end{eqnarray}
\noindent
Here
\begin{equation}
{\mathcal M}\left({\bf x},H, \zeta \right)=\int\frac{d^2{\bf q}_{\perp}}{\left(2 \pi\right)^2}\;e^{i{\bf q}_{\perp}\cdot{\bf x}}
\int_{1}^{\infty} d p \frac{2p^4-2p^2+1}{\left[4p^2+\left(c q_{\perp}/\zeta\right)^2\right]^{3/2}}
\;e^{-\frac{\zeta}{c}H\sqrt{4p^2+\left(c q_{\perp}/\zeta\right)^2}   },\label{FinalM4}
\end{equation}
\noindent
\begin{equation}
 \overline{\delta \epsilon}(i \zeta)= \frac{{\delta \epsilon(i \zeta)}}{{1+\frac{1}{3} \delta
\epsilon(i \zeta)  }},
\end{equation}
\noindent
${\bf q}_{\perp}=(q_x,q_y)$, and $q_{\perp}=\sqrt{q_x^2+ q_y^2}$.

Note that in the case of metals, the dielectric contrast $\delta \epsilon\left(i \zeta\right)=\epsilon\left(i \zeta\right)-1$ is large, but
$ \overline{\delta \epsilon}(i \zeta)= \frac{{\delta \epsilon(i \zeta)}}{{1+\frac{1}{3} \delta\epsilon(i \zeta)  }}= 3 \frac{\epsilon\left(i \zeta\right)-1}{\epsilon\left(i \zeta\right)+2}$ is not large. In case of two {\it flat} and parallel perfect conductors,
$\epsilon\left(i \zeta\right)\rightarrow \infty$ and Eq. (2) yields $E_C= -\frac{69}{640 \pi^2}  \frac{\hbar c}{H^3}$ which is smaller than the exact
result $ -\frac{\pi^2}{720 }\frac{\hbar c}{H^3} $ by a factor about $0.8$\ \cite{Ramin00}.
We expect DCP to exhibit the same order of error in the case of corrugated metals.

We now focus on periodic uniaxial corrugations, where $h_{1,2}({\bf x})=h_{1,2}({x})$ and $h_{1,2}({x}+\lambda)=h_{1,2}({x})$.
It is convenient to recast the kernel ${\mathcal M}\left({\bf x},H, \zeta \right)$ in a different form.
First, we use the identity
\begin{equation}
e^{-\frac{\zeta H}{c}\sqrt{4p^2+\left(c q_{\perp}/\zeta\right)^2}} \!\!=\!\!
\int \! \frac{dQ_{z}}{ \pi}  \frac{\sqrt{4p^2+\left(c q_{\perp}/\zeta\right)^2}}{ 4p^2+\left(c q_{\perp}/\zeta\right)^2+ Q_{z}^2   } e^{i \frac{\zeta}{c}H Q_{z} },
\end{equation}
\noindent
and then employ
\begin{eqnarray}
\frac{1}{4p^2  \!\!+\!\!\left(c q_{\perp}/\zeta\right)^2}  \frac{1}{4p^2 \!\!+\!\!\left(c q_{\perp}/\zeta\right)^2 \!\!+\!\! Q_{z}^2   }
&= & \!\! \int_0^\infty \!dt_1 \!\! \int_0^\infty \! dt_2 \! \exp[ -\!(t_1\!+\!t_2)(4p^2 \!+\!c^2 q_{\perp}^2/\zeta^2)\!-\!t_2  Q_{z}^2]  ,\nonumber \\
&= &\!\!\int_0^1 \! dt \! \int_0^\infty ds \;s \exp[-s(4p^2+c^2 q_{\perp}^2/\zeta^2)-st Q_{z}^2 ].
\end{eqnarray}
\noindent
This representation of the kernel allows us to write
\begin{eqnarray}
E_{C}   &=& -\frac{\hbar L_{y}}{ 8 \pi^3   c^3}
\int_{1}^{\infty}dp \int_{0}^{\infty}ds \int_{0}^{1}dt    \left(2p^4 - 2 p^2 +1\right)  t^{-1/2} \\
& & \! \! \! \! \! \times \int_{0}^{\infty}d\zeta \; \int dx\, dx^{\prime} \; \zeta^3\,\overline{\delta\epsilon_{1}}(i \zeta) \, \overline{\delta\epsilon_{2}}(i \zeta)\;
 e^{-\left(\zeta/c\right)^2 \frac{(x-x^{\prime})^2}{4s}-\frac{\left(\zeta/c\right)^2}{4 s t}\,
\left[H+h_{2}\left(x\right)-h_{1}(x^{\prime})\right]^2 -\, 4 p^2 s    }.\nonumber
\label{EnergyLat44}
\end{eqnarray}
Here $L_{x}$ and $L_{y}$ are the extensions of the system in the $x$ and $y$ directions, respectively.
$A=L_{x} L_{y}$ is the area of each plate.

The Casimir energy strongly depends on the geometrical features, especially the relative lateral
displacement of the plates $b \lambda $; see Fig. (\ref{fig:sifig}). The plates thus experience
the lateral Casimir force
\begin{equation}
F^{\rm lat}=-\frac{1}{\lambda}\frac{\partial E_{C}}{\partial b}\label{lateraleq4}.
\end{equation}

\section{Results}

We study gold and silicon plates. We use the plasma model
\begin{equation}
\epsilon (i\zeta)=1+\frac{\omega_{p}^{2}}{\zeta^{2}} \label{plasma}
\end{equation}
with $\omega_{p}=1.3 \times 10^{16}\;{\rm rad/s} $ to model the permittivity of gold.
We use the Drude-Lorentz model
\begin{equation}
\epsilon (i\zeta)=\epsilon_h+\frac{( \epsilon_l - \epsilon_h)\omega_{0}^{2}}{\zeta^{2}+\omega_{0}^{2} }
\end{equation}
with $\epsilon_h=1.035$, $\epsilon_l=11.87$, and $\omega_{0}= 6.6\times 10^{ 15}\;{\rm rad/s} $
to model the dielectric function of silicon; c.f. Ref.\ \cite{lam}.
\noindent
In Eq. (\ref{EnergyLat44}), we  use the identity
\begin{equation}
\int_{-\infty}^{\infty}{\rm d}x\,f(x)= \sum_{n=-\infty}^{\infty}\int_{0}^{\lambda}{\rm d}x\,f(x-n \lambda),
\end{equation}
\noindent
which is valid for any function $f(x)$ with periodicity $\lambda$, to simplify the numerical evaluation of $E_{C}$.

\begin{figure*}[h]
\includegraphics[width=.45\textwidth]{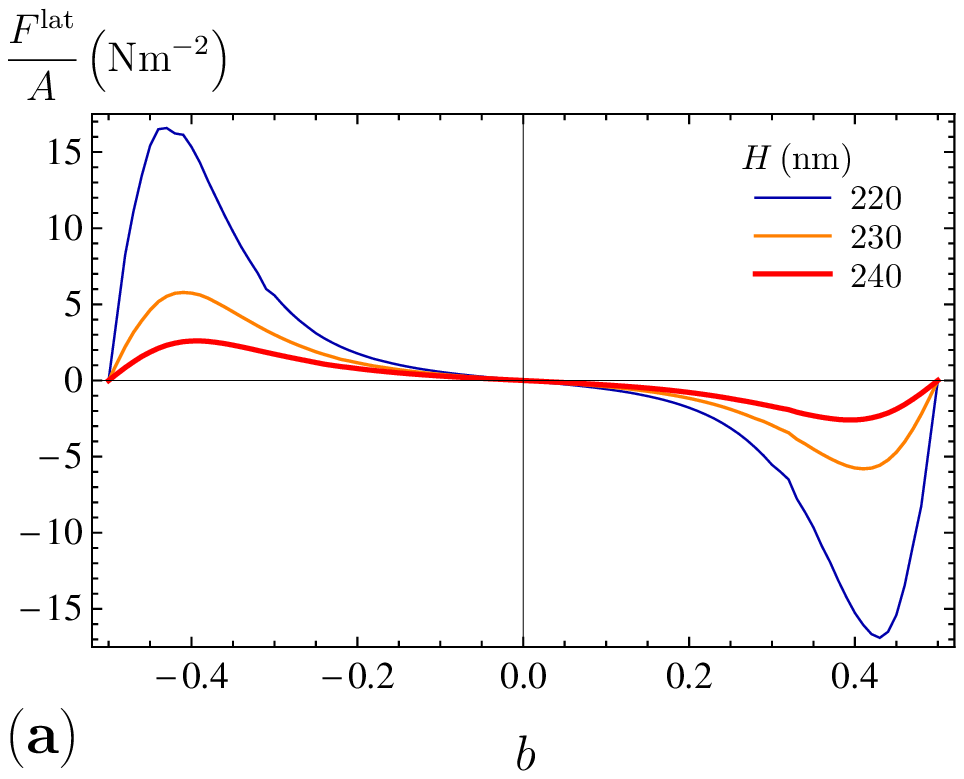}
\hspace{.5cm}
\includegraphics[width=.45\textwidth]{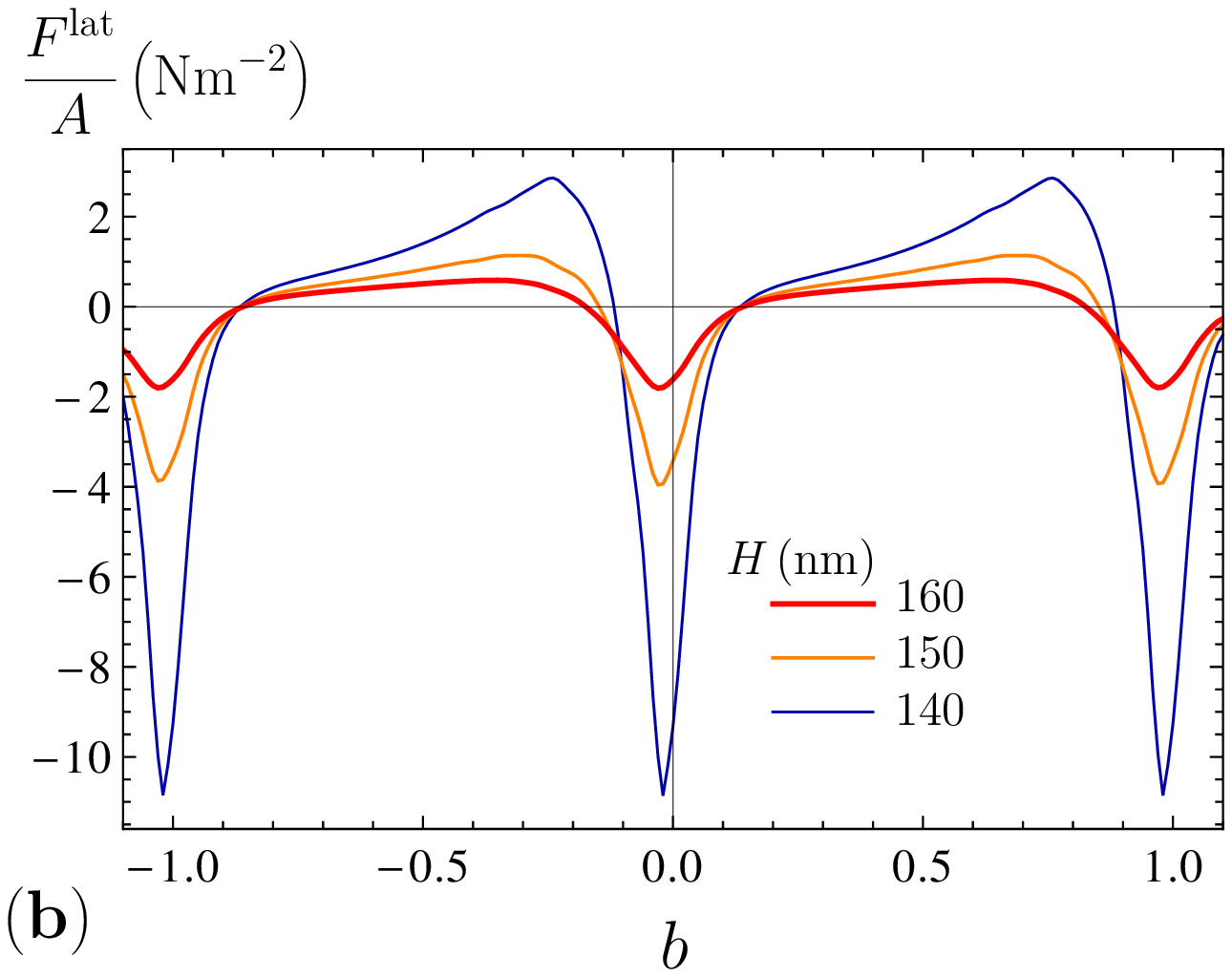}
\vspace{.5cm} \\
\includegraphics[width=.45\textwidth]{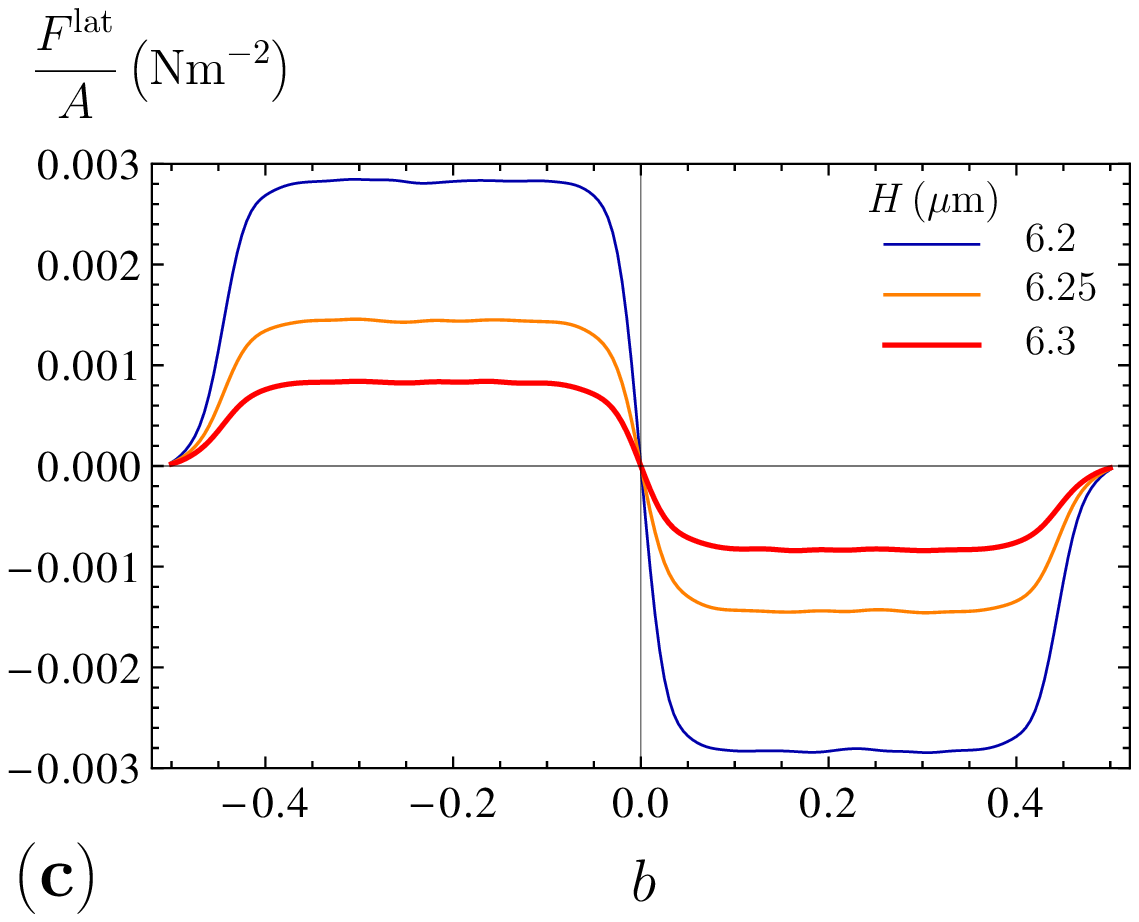}
\hspace{.5cm}
\includegraphics[width=.44\textwidth]{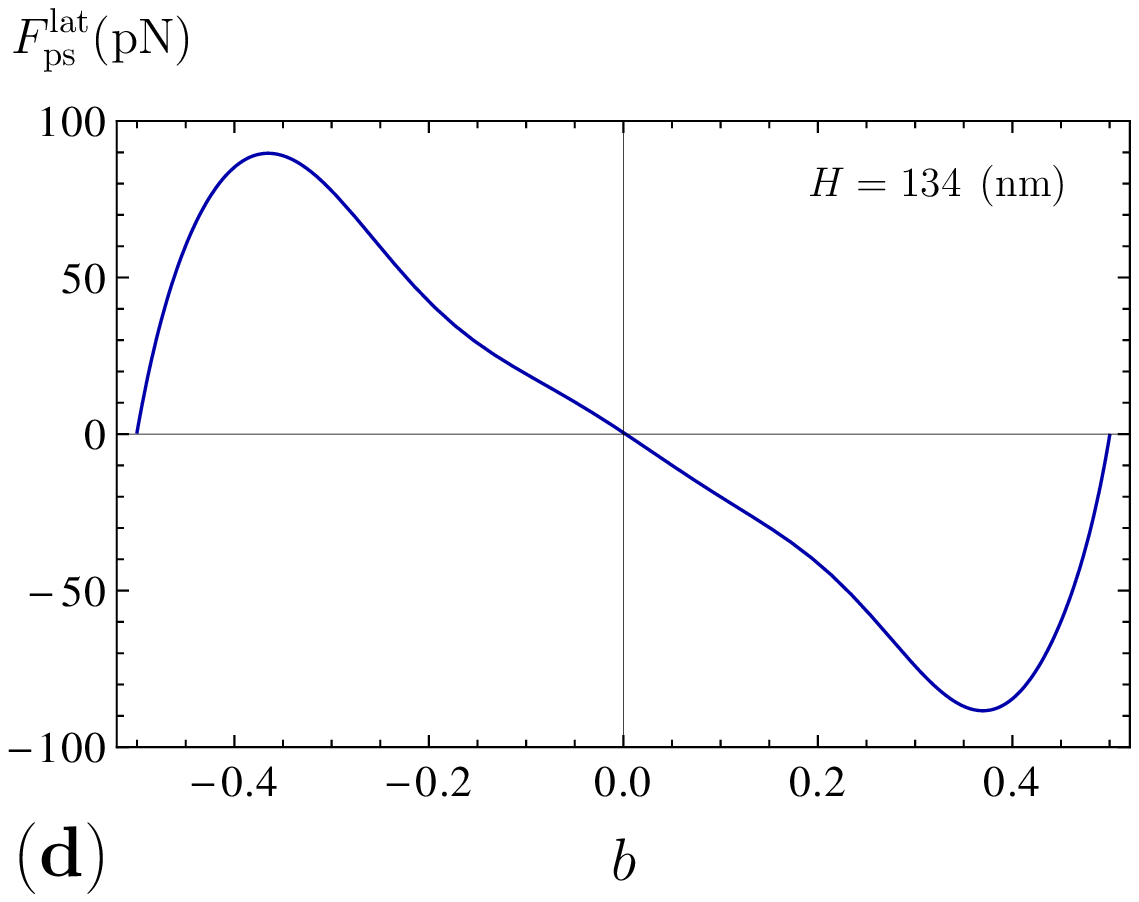}
\caption{ (Color online) $F^{\rm lat}/A$ as a function of dimensionless parameter $b$ for various $H$, for the case of: (a) Deep sinusoidal corrugations on gold plates.
Here $a_{1}= 120\; {\rm nm}$, $a_{2}= 80\; {\rm nm}$ and $\lambda= 500 \;{\rm nm}$. (b) Deep sawtooth
corrugations on gold plates. Here $a_{1}= 55\; {\rm nm}$, $a_{2}= 70\; {\rm nm}$ and $\lambda= 530 \;{\rm nm}$. (c) Deep rectangular
corrugations on silicon plates. Here $f_{1}=f_{2}= \frac{5}{9}$, $a_{1}=a_{2}= 6 \; {\rm \mu m}$, and $\lambda= 9 \;{\rm \mu m}$. (d) $F^{\rm lat}_{\rm ps}$ as a function of dimensionless parameter $b$. Here $H=134\; {\rm {nm}}$,
$a_{1}= 85.4\; {\rm nm}$, $a_{2}= 25.5\; {\rm nm}$, $\lambda= 574.7 \;{\rm nm}$,
and $2R=194 \;{\rm \mu m} $.
} \label{fig:all_fig}
\end{figure*}

First, we study deep sinusoidal corrugations for gold plates; see Fig. \ref{fig:sifig}. The height functions are
\begin{eqnarray}
&&h_{1}(x)=a_{1}\,\sin \left(\frac{2 \pi}{\lambda}x\right) ,\nonumber \\
&&h_{2}(x)=a_{2}\,\sin \left(\frac{2 \pi}{\lambda}x+b\right).
\end{eqnarray}
We assume $a_{1}= 120\; {\rm nm}$, $a_{2}= 80\; {\rm nm}$, and $\lambda= 500 \;{\rm nm}$.
Figure\ \ref{fig:all_fig}(a) demonstrates $F^{\rm lat}$ as a function of $b$ for various values of $H$.
As expected, $F^{\rm lat}$ is an odd function of $b$. $F^{\rm lat}$ decays rapidly as $H$ increases.
$F^{\rm lat}$ is zero at $b=0$ and $b=\pm 0.5$. Interestingly, $F^{\rm lat}$ is almost $b$-independent in the range $ -0.1< b< 0.1$. For $H=220 \;{\rm nm} $, $230 \;{\rm nm} $, and
$240 \;{\rm nm} $, $F^{\rm lat}$ gains its maximum at $b=\pm 0.43$, $\pm 0.41$, and $\pm 0.39$, respectively. For $H=220 \;{\rm nm} $, the maximum of
$F^{\rm lat}/A$ is greater than $15 \;{\rm N}{\rm m}^{-2}$, which is a considerable pressure.
Note that $F^{\rm lat}(b)$ for deep sinusoidal corrugations on gold, is quite distinct from $ F^{\rm lat}_{\rm{max}} \sin(2 \pi b) $ which is expected
for shallow sinusoidal corrugations on perfect reflectors\ \cite{Emig2001}. In case of shallow corrugations,
$ F^{\rm lat}_{\rm{max}} \propto a_1 a_2 $, whereas for deep corrugation $F^{\rm lat}_{\rm{max}}$ does not depend linearly on the corrugation amplitudes.

To emphasize the influence of geometry on the lateral Casimir force, we consider deep sawtooth corrugations
on gold plates.
We assume $a_{1}= 55\; {\rm nm}$, $a_{2}= 70\; {\rm nm}$ and $\lambda= 530 \;{\rm nm}$.
Figure\ \ref{fig:all_fig}(b) demonstrates $F^{\rm lat}$ as a function of $b$ for various values of $H$.
Due to the geometry of the system, $F^{\rm lat}$ is nonzero at $b=0$ and $b=\pm 1$.
A notable feature of this case is the rapid change of $F^{\rm lat}$ around $b=0$ and $b=\pm 1$.
$F^{\rm lat}$ is neither an odd nor an even function of $b$. In both $b>0$ and $b<0$ regions, $F^{\rm lat}$ gains both positive and negative values. For each distance $H$, $F^{\rm lat}$ has two zeroes
in the $ 1>b>0$ ($-1<b<0$) region. For example, the lateral Casimir force is zero at
$b=0.14$ and $b=0.89$, if $H= 140\;{\rm nm}$.

Next, we consider deep rectangular corrugations on silicon plates. We assume $f_{1}=f_{2}= \frac{5}{9}$,
$a_{1}=a_{2}= 6 \; {\rm \mu m}$, and $\lambda= 9 \;{\rm \mu m}$; see Fig. \ref{fig:sifig}. We study the lateral force for mean separations $H> 6 \;{\rm \mu m}$. These values are relevant to the recently fabricated system \cite{ema}. Figure\ \ref{fig:all_fig}(c) demonstrates $F^{\rm lat}$ as a function of $b$ for various values of $H$. Here the maximum of $F^{\rm lat}/A$ is
about $0.003\;{\rm N}{\rm m}^{-2}$. Note that in contrast to the previous examples, here the typical length scales are of the order of a few microns. As one would expect from geometrical considerations,
$F^{\rm lat}$ is an odd function of $b$. $F^{\rm lat}$ is zero at $b=0$ and $b=\pm 0.5$. We note that $F^{\rm lat}$ has an almost constant value in the region $ 0.05<|b|<0.4$.

It is instructive to compare the results of DCP and other approximate methods.
We compute the lateral Casimir force $ F_{\rm ps}^{\rm lat} $ between sinusoidally corrugated surfaces of a plate and a sphere of radius $R$.
We consider the experimentally realized parameters
$a_{1}= 85.4\; {\rm nm}$, $a_{2}= 25.5\; {\rm nm}$, $\lambda= 574.7 \;{\rm nm} $, $H=134 \; {\rm {nm}}$,
and $2R=194 \;{\rm \mu m} $\ \cite{chiu}. According to the PFA
\begin{equation}
F_{\rm ps}^{\rm lat}=\frac{2 \pi R}{\lambda} \;\frac{\partial}{\partial
b}\int_{H}^{\infty} d H' \; \frac{E_{C}(H')}{A}.
\label{p-s}
\end{equation}
To model the permittivity of gold, we use the generalized plasmalike model of Ref.\ \cite{chiu}.
Figure \ref{fig:all_fig}(d) demonstrates $F^{\rm lat}_{\rm ps}$ as a function of $b$. $F^{\rm lat}_{\rm ps}$
is a nonsinusoidal function of $b$ and gains its maximum $ 89.7  \; {\rm pN}$ at $b= \pm 0.36 $.
Using scattering theory and PFA, Chiu {\it et al.} found that $F^{\rm lat}_{\rm ps}$ deviates from a sinusoidal form and
reaches its maximum $ 57 \; {\rm pN}$ at $b=\pm 0.34$. Experimental value of the maximum force is $ 52.5 \pm 12  \; {\rm pN}$.

\subsection{Discussions}

As mentioned before, {approximate} methods are quite useful in early stages of nanomachine design.
One can obtain the Casimir energy of two corrugated surfaces using the PWS of potential $U(r)= - B \hbar c/ r^7$.
The factor $B$ is chosen to recover the exact Casimir energy of two flat plates\ \cite{Klim-etal-RMP09}. The result of PWS is not valid if  $ \lambda < H$\ \cite{Emig2001}. PFA also fails in the limit $ \lambda < H$\ \cite{over1}. DCP allows us to consider arbitrary dielectric functions and
arbitrary deformations with {large} amplitudes. However, even using the Clausius-Mossotti summation of terms,
DCP underestimates the Casimir force by a factor about $0.8$.
Lifshitz and numerical results agree well as one utilizes the sixth order term of DCP\ \cite{Ramin00}.
Thus DCP offers a {systematic} inclusion of many-body fluctuation-induced interactions, which are overlooked by PWS and PFA.

In summary, we use the Clausius-Mossotti summation of the dielectric contrast perturbation theory to calculate
the lateral Casimir force between deeply corrugated dielectric plates. Figures\ \ref{fig:all_fig}(a)-(c) demonstrate clearly that one can employ
various geometries and materials to engineer the displacement-dependence of the lateral Casimir force.
Our work lends itself to the proposition that one can harness such non-sinusoidal lateral forces to design mechanical rectifiers and sensors.

\section*{Acknowledgements}

This work was supported by EPSRC under Grant EP/F036167/2.

\end{document}